\documentclass[twocolumn]{svjour3}          
\smartqed  
\usepackage{graphicx}
\usepackage{latexsym}
\usepackage{url}
\usepackage{amssymb}
\usepackage{amsmath}
\usepackage{graphicx}
\usepackage{epsfig}
\usepackage{graphics}
\usepackage[latin1]{inputenc}
\usepackage[numbers,sort]{natbib}
\usepackage[normalem]{ulem}
\journalname{General Relativity and Gravitation}

\begin{document}

\title{
  Structure of Compact Stars in $R$-squared Palatini Gravity 
}


\author{Florencia~A.~Teppa~Pannia$^{1,*}$\and~Federico~Garc\'{\i}a$^{1,2,*}$\and~Santiago~E.~Perez~Bergliaffa$^{3}$\and~Mariana~Orellana$^{4,**}$\and~Gustavo~E.~Romero$^{1,2,**}$}

\authorrunning{Teppa Pannia et al.} 

\institute{ \at $^{*}$ Fellow of CONICET. \at $^{**}$ Member of CONICET. \at $^{1}$ Facultad de Ciencias Astron\'omicas y Geof\'{\i}sicas, Universidad Nacional de La Plata, Paseo del Bosque s/n, 1900 La Plata, Argentina. \at $^{2}$ Instituto Argentino de Radioastronom\'{\i}a CCT La Plata (CONICET), C.C.5 (1894) Villa Elisa, Buenos Aires, Argentina. 
\at $^{3}$  Departamento de F\'isica Te\'orica, Instituto de F\'isica, Universidade do Estado de Rio de Janeiro, Brasil.
\at $^{4}$ Sede Andina de la Universidad Nacional de R\'io Negro, Argentina. \\
\email{fteppa@fcaglp.unlp.edu.ar}
}

\date{Received: date / Accepted: date}

\maketitle

\begin{abstract}
  We analyse configurations of neutron stars in the so-called $R$-squared gravity in the Palatini formalism.
Using a realistic equation of state we show that the mass-radius configurations are lighter than their counterparts in General Relativity. We also obtain  the internal profiles, which run in strong correlation with the derivatives of the equation of state, leading to regions where the  mass parameter decreases with the  radial coordinate in a counter-intuitive way. 
In order to analyse such correlation, we introduce a parametrisation of the equation of state given by multiple polytropes, which allows us to explicitly control its derivatives. We show that, even in a limiting case where hard phase transitions in matter are allowed, the internal profile of the mass parameter still presents strange features and the calculated $M-R$ configurations also yield neutron stars lighter than those obtained in General Relativity.  

\end{abstract}

\section{Introduction}
\label{intro}
The so-called Extended Theories of Gravity (ETGs) are generalisations of General Relativity (GR) conceived to deal with theoretical and observational issues arising from astrophysical and cosmological scenarios (see \cite{Capozziello2011} for an extended review).
A particular class of them, namely $f(R)$ theories, is obtained by substituting the Einstein-Hilbert Lagrangian density by a function of the Ricci scalar curvature $R$.

In the low-curvature regime, one of the stronger motivations to study $f(R)$ theories is to describe cosmological observations without the necessity of invoking a dark energy component in the current epoch of the evolution of the universe \cite{Sotiriou2006,Sotiriou2010,deFelice2010,Capozziello2011,Nojiri2011}. In this vein, there are several $f(R)$ models that successfully account for the succession of different cosmological eras, and satisfy the current Solar System and laboratory constraints \cite{Nojiri2006,Starobinsky2007,HuSawicki2007,Cognola2008,Miranda2009,Jaime2011}.

A different motivation to consider $f(R)$ theories comes from the fact that the scarce data available from phenomena in the strong-curvature regime are compatible not only with GR, but also with $f(R)$ and other modified theories (see for instance \cite{Konoplya2016,Vainio2016}). In this context, Neutron Stars (NSs) may offer the possibility of testing deviations from GR through astrophysical observations. The internal structure of such compact objects is described in GR by the solutions of the Tolman-Oppenheimer-Volkoff (TOV) equations, together with a suitable Equation of State (EoS). In the framework of $f(R)$-theories in the metric formalism \cite{deFelice2010}, the internal structure of NSs has been previously studied by several authors. Since the modified TOV equations have derivatives of the metric up to the fourth order, different approaches have been developed to deal with the numerical integration. 
 One of the first attempts was to consider the solution inside the star as a perturbation of the GR case, and match it with the external solution characterised by the Schwarzschild metric. This perturbative method was used in \cite{Cooney2010,Arapoglu2011,Orellana2013,Astashenok2013} to analyse the internal structure of NSs using polytropic and realistic EoSs to describe the matter content inside the compact object. The structure of NSs using a perturbative approach and including hyperons and/or quarks EoSs was also explored in \cite{Astashenok2014}. 

 However, as it was pointed out in \cite{Yazadjiev2014}, the use of a perturbative method to investigate the strong field regime in $f(R)$ theories and may lead to unphysical results. Self-consistent models of NSs are then required to solve simultaneously for the internal and external regions, assuming appropriate boundary conditions at the centre of the star and at infinity. This new approach was explored by introducing a scalar field and working in the so-called Jordan frame \cite{Yazadjiev2014}, by recasting the field equations without mapping the original $f(R)$ theory to any scalar-tensor counterpart \cite{Salgado2011}, and by using self-consistent numerical methods to solve simultaneously the internal structure of the star and the external metric \cite{Astasheno2015b,Capozziello2016,Alvaro2016}.\footnote{More sophisticated models of NSs were also considered in the framework of $f(R)$, such those including rotation \cite{Staykov2014,Yazadjiev2015} and strong magnetic mean fields \cite{Astashenok2015}.}

The internal structure of NSs has also been studied using the Palatini formalism, in which the metric and the connection are a priori considered as independent geometrical entities \cite{Olmo2008,Olmo2011}. This approach has the advantage of straightforwardly yielding field equations  with derivatives of the metric up to second order. The modified TOV equations in this case were firstly derived in \cite{Kainulainen2007} by matching the interior solution with the exterior Schwarzschild-de Sitter solution. 
 Let us remark that, differently from  the above-mentioned metric approach, in Palatini gravity the unique solution of static and spherically symmetric vacuum configurations is the Schwarzschild-de Sitter metric, in which the value of the effective cosmological constant  is calculated using the well-known equivalence of $f(R)$ and  Brans-Dicke theories 
(see for instance \cite{Capozziello2011}). In the case of a null cosmological constant (which is precisely that of $R$-squared gravity), the mass parameter coincides with the Schwarzschild mass, (\emph{i.e.} with the value of $m(r)$ at the surface of the star.  

The structure of static and spherically-symmetric compact stars in the context of the Palatini formalism was studied in \cite{Barausse2008a} assuming both polytropic and realistic EoSs. In the first case, the authors showed that the matching between the interior and  exterior solutions at the surface of the star can yield divergences in the curvature invariants near the surface of the star when polytropic EoSs with $3/2< \Gamma < 2$ are used for generic $f(R)$.

The no-go theorem related to the issue of the singularity at the surface of some polytropic NSs in these model was carefully analysed in \cite{Barausse2008c}. It was claimed there that the origin of the singularity does not lie in the fluid approximation or in the specifics of the approach followed to solve the internal structure of the star, but is related to the intrinsic features of Palatini $f(R)$ gravity. The authors of \cite{Barausse2008c} argued that the root of the problem lies in the differential structure of the field equations, in which the matter field derivatives are of higher order than the metric derivatives.\footnote{In fact, this feature induces corrections to the standard model of particle physics at the MeV energy scale, see \cite{Flanagan2004,Flanagan2004b,Iglesias2007}.}   This peculiarity introduces non-cumulative effects and makes the metric sensitive to the local characteristics of matter. A possible resolution to the singularity problem in this context, namely the addition of terms 
quadratic in the derivatives of the connection
to the gravitational action, 
was also discussed in \cite{Barausse2008c}.\footnote{The existence of singularities at the surface of the star in the context of Eddington-inspired Born-Infeld (EiBI) theory was proved in \cite{Pani2012}, while a possible resolution to this problem, due to gravitational back-reaction on the particles was presented in \cite{Kim2014}.}$^{,}$\footnote{It was shown in \cite{Olmo2008} that the surface singularities are not physical in the case of Planck-scale modified Lagrangians, in which they are instead an artifact of the idealised equation of state used.} 

If more realistic EoSs (which take into account the fundamental microphysics of the matter that composes the star) are used along with the modified TOV equations, compact stars present another unappealing feature in Palatini $f(R)$ gravity. In \cite{Barausse2008a}, the structure of NSs was calculated for the choice $f(R)=R+\alpha R^2$, using an analytic approximation of the realistic FPS EoS \cite{Haensel2004}. In spite of the fact that such EoS yields a regular solution at the surface, the interior metric strongly depends on the first and second derivatives of the function $\rho(p)$. As a consequence, the radial profiles of the mass parameter are not smooth functions as in GR, but develop bumps when there are rapid changes in the derivatives of the EoS \cite{Barausse2008c}.

The above results show that the modelling of NSs in $f(R)$ theories in the Palatini formalism involves some extra considerations when compared with the GR case, due to the strong correlation between the metric and the derivatives of the EoS. It is important to note that these are poorly constrained, since the EoSs are actually constructed to fit only the zeroth-order relation between $\rho$ and $p$, which is enough to calculate the structure of NSs in GR. Thus, special care must be taken if high-order derivatives (e.g. ${\rm d}p/{\rm d}\rho$, ${\rm d}^2p/{\rm d}\rho^2$) are used during the calculation, as in the case we are interested in here.

 The main goal of this work is to check whether the non-smoothness of the mass parameter reported in \cite{Barausse2008a} is actually a feature of $f(R)$ theories in the Palatini formalism or it may be due to the details of the EoS chosen there. For this purpose, 
 we calculate the structure of a star in the Palatini formalism with the choice $f(R)=R+\alpha R^{2}$ in two different ways. First, we use the SLY EoS (instead of the FPS EoS used in \cite{Barausse2008a}). As a second test, an approximation to the EoS based on the connection of multiple polytropes was employed. The polytropes represent the state of matter at the core and crust of the NS, and allow us to control the derivatives of the EoS through a set of parameters. 
We find that in both cases the internal profiles run in strong correlation with the derivatives of the EoS, leading to regions where the mass parameter decreases with the radial coordinate in a counter-intuitive way, even in the case where hard phase transitions in the EoS are allowed. We also find that mass-radius configurations in this theory do not allow heavier NSs than in GR for any plausible $\alpha > 0$.

The paper is organised as follows. In Section~2, we present the modified TOV equations in the Palatini formalism. Realistic EoSs and the integration of the stellar structure are described in Section~3, focusing on the mass-radius relations and the correlation between the features of the internal profile and the first and second derivatives of the EoSs. In Section~4 we introduce a parametrisation for the EoS based on the connection of multiple polytropes, and examine the stellar structure obtained for this EoS. Final remarks are presented in Section~5.

\section{Stellar structure in \emph{f(R)} Palatini gravity}
\label{sec:1}
The modified Hilbert-Einstein action is given by
\begin{equation}
S[g_{\mu\nu},\Gamma,\psi_{\rm m}]=\frac{c^4}{16 \pi G}\int{{\rm d}^4x \sqrt{-g} f(R)} + S_{\rm m}[g_{\mu\nu},\psi_{\rm m}], 
\end{equation}
\noindent where $f(R)$ is a function of the Ricci scalar $R\equiv g^{\mu\nu}R_{\mu\nu}(\Gamma)$, with $R_{\mu\nu}(\Gamma)=-\partial_\mu\Gamma^{\lambda}_{\lambda\nu}+ \partial_\lambda\Gamma^{\lambda}_{\mu\nu}+ \Gamma^\lambda_{\mu\rho}\Gamma^{\rho}_{\nu\lambda}- \Gamma^{\lambda}_{\nu\rho}\Gamma^{\rho}_{\mu\lambda}$. 
 The matter action $S_{\rm m}$ depends on the matter fields $\psi_{\rm m}$ and the metric $g_{\mu\nu}$.

 In the Palatini formalism the field equations are obtained by varying the action with respect to the metric and the connection \cite{Olmo2008}, and they are given by 
\begin{equation}
  \label{wrtg}
 f_R(R)R_{\mu\nu}(\Gamma)-\frac{1}{2}f(R) g_{\mu\nu} = \frac{8\pi G}{c^4}T_{\mu\nu}\,, 
\end{equation}
\begin{equation}
  \label{wrtgamma}
  \nabla_{\rho}\left[\sqrt{-g}\left(\delta^{\rho}_{\lambda}f_R g^{\mu\nu} -\frac{1}{2}\delta^{\mu}_{\lambda}fRg^{\rho\nu}-\frac{1}{2}\delta^{\nu}_{\lambda}f_Rg^{\mu\rho} \right) \right]=0\,,
\end{equation}
where $f_R\equiv {\rm d}f/{\rm d}R$ and $T_{\mu\nu}$ is the energy-momentum tensor, which satisfies the continuity equation 
\begin{equation}
\label{conteq}
\nabla_{\mu}T^{\mu\nu}=0.
\end{equation}
 The trace of Eqn.~(\ref{wrtg}) yields 
\begin{equation}
  \label{trace}
f_R(R)R-2f(R)=\frac{8\pi G}{c^4}T.
\end{equation}
This algebraic equation can be used to express the scalar curvature $R$ as a function of the trace $T$ of the energy-momentum tensor. 

The stellar structure is computed by assuming a spherically-symmetric and static metric with line element
\begin{equation}
{\rm d}s^2=-e^{A(r)}c^2{\rm d}t^2+e^{B(r)}{\rm d}r^2+r^2({\rm d}\theta^2+\sin^2{\theta}{\rm d}\phi^2),
\end{equation}
and a perfect-fluid matter with energy-momentum tensor $T_{\mu\nu}=(c^2\rho + p)u_\mu u_\nu + p g_{\mu\nu}$, where $\rho(r)$ is the density and $p(r)$ is the pressure.
 With these considerations, the continuity equation (\ref{conteq}) yields  
\begin{equation}
  p'=-\frac{A'}{2}(c^2\rho+p)\, ,
\end{equation}
and the $tt$ and $rr$ components of the field equations (\ref{wrtg}) can be written as \cite{Barausse2008a,Reijonen2009}
\begin{eqnarray}
A'&=&-\frac{1}{1+\gamma_0} \left( \frac{1-e^{B}}{r} - \frac{e^{B}}{f_R} \frac{8 \pi G r p}{c^4} + \frac{\alpha_0}{r}  \right)\,, \\
\label{dAdr}
B'&=& \frac{1}{1+\gamma_0} \left( \frac{1-e^{B}}{r} + \frac{e^{B}}{f_R}\frac{8 \pi G r \rho}{c^2}   + \frac{\alpha_0 + \beta_0}{r} \right) ,
  \label{dBdr} 
\end{eqnarray}
where prime symbol denotes derivative with respect to the radial coordinate, $r$, and 
\begin{eqnarray}
\alpha_0 &\equiv& r^2 \left( \frac{3}{4} \left( \frac{f_R'}{f_R} \right)^2 + \frac{2 f_R'}{ r f_R} + \frac{e^B}{2} \left(R- \frac{f}{f_R}\right) \right) , \\
\beta_0 &\equiv&  r^2 \left(\frac{f_R''}{f_R} -\frac{3}{2} \left(\frac{f_R'}{f_R} \right)^2\right), \\
\gamma_0 &\equiv& \frac{r f_R' }{2f_R}.
\end{eqnarray}
 The generalised TOV equations take the form \cite{Kainulainen2007,Reijonen2009}
\begin{eqnarray}
  &&p'=-\frac{1}{1+\gamma_0}\ \frac{c^2\rho + p}{r(c^2r-2G m)} \nonumber \\
   && \qquad \qquad \left(Gm+\frac{4 \pi G r^3 p}{f_R} - \frac{\alpha_0}{2} (c^2r - 2G m) \right)\,,
\label{dpdr}  \\
&&m'=\frac{1}{1+\gamma_0} \left( \frac{4 \pi r^2 \rho}{f_R} + \frac{c^2}{G}\frac{\alpha_0 + \beta_0}{2}  \right. \nonumber \\
 && \qquad \qquad \qquad \qquad \qquad \quad \left. -\frac{m}{r}(\alpha_0 + \beta_0- \gamma_0) \right)\,, 
\label{dmdr} 
\end{eqnarray}
\noindent where the mass parameter is defined as $m(r)\equiv c^2r(1-e^{-B})/2G$.\footnote{An alternative representation of the TOV equations can be found in \cite{Olmo2011}.}. 
From now on, we will work with a particular form for the $f(R)$, the so-called $R$-squared gravity, characterised by the function $f(R)=R+\alpha R^2$.\footnote{Negative powers of $R$ are negligible in the strong field regime in which we are interested in here \cite{Sotiriou2006}.}
 The constant $\alpha$ is a free parameter of the theory which must be positive due to stability considerations \cite{deFelice2010,Sotiriou2010}. This type of theory has been frequently studied due to its renormalisation properties \cite{Stelle1977}. In a cosmological context, it was shown in \cite{Starobinsky1980} that this theory gives rise, in the metric formalism, to an early non-singular period of accelerating expansion. Also in the metric formalism, slowly-rotating NSs were analysed in \cite{Staykov2014}, and  
the behavior of the normalized I-Q relation for neutron stars was discussed in \cite{Doneva2015}. Charged black holes in $f(R) = R+\alpha R^2$ in Palatini formalism have been analysed in \cite{Olmo2011b}, and
the ratio of crustal to the total moment of inertia of NSs in these theories was calculated in \cite{Staykov2015}. Bouncing cosmologies in the Palatini version of this theory have been studied in \cite{Barragan2009}.

The system of differential equations (\ref{dpdr})-(\ref{dmdr}) can be solved if a relation between $\rho$ and $p$ is given. Note that using Eqn.~(\ref{trace}) the scalar curvature $R$ can be expressed as a function of $T$. 
In particular, $R$-squared gravity yields $R=-8 \pi G T /c^4 =-8 \pi G (-c^2\rho + 3p)/c^4$.  
Through the chain rule, the derivatives of $f_R$ with respect to the radial coordinate, $r$, in the functions $\alpha_0$, $\beta_0$ and $\gamma_0$, are written in terms of $p'$, $p''$, and the first and second derivatives of the EoS. 
 Then, the calculation of the stellar structure requires a non-trivial derivation of Eqns.~(\ref{dpdr}) and (\ref{dmdr}) in an explicit form \cite{Reijonen2009}. 

\section{$M-R$ configurations and internal profiles }

\subsection{Equation of State}
The EoS contains the information of the behaviour of matter inside NSs through several orders of magnitude in density. Because the properties of  matter at the highest densities in the central region of NSs are not well understood, different EoSs have been proposed and constrained with observations of masses and radii of actual NSs \cite{Douchin2001,Haensel2004}.

It is important to emphasise that the EoSs available in the literature are usually given by the tabulation of the zeroth-order relation between $\rho$ and $p$, because such is the relation needed to calculate the structure of NSs in GR. However, in such cases the usual interpolation technique fails to accurately represent high-order derivatives \cite{Eksi2014}.

\begin{figure}
\begin{center}
\resizebox{\hsize}{!}{\includegraphics[angle=-90]{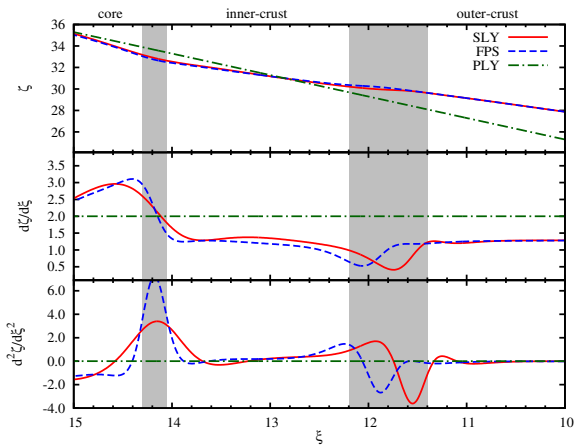}}
\caption{Analytic approximation of SLY and FPS EoSs (upper panel), and their corresponding first and second derivatives (middle and bottom panel, respectively). The simpler polytropic approximation PLY is also included for comparison. Grey-shadowed areas indicate transition regions.}
\label{EOS_SLY_FPS_PLY}
\end{center}
\end{figure}

 Thus, special care should be taken if ${\rm d}p/{\rm d}\rho$ and ${\rm d}^2p/{\rm d}\rho^2$ are used during the calculation, as in the case we are interested in here. Alternatively,  analytic approximations instead of tabular EoSs can be used in order to achieve more precision. In this direction, we first consider the SLY EoS, extensively used to calculate the internal structure of NSs \cite{Douchin2001,Haensel2004}, as well as the FPS EoS, used in \cite{Barausse2008a} for comparison. They are complex representations of tabular EoSs obtained through a thermodynamically-consistent procedure to best-fit coefficients of a polynomial expansion, both in the crust and core density regimes \cite{Haensel2004}. The analytic parametrisations for SLY and FPS EoSs are given by
\begin{eqnarray}
  \zeta&=&\frac{a_1+a_2\xi+a_3\xi^3}{1+a_4\,\xi}\,f_0(a_5(\xi-a_6)) \nonumber  \\
    && \quad + (a_7+a_8\xi)\,f_0(a_9(a_{10}-\xi))\nonumber  \\
    && \quad \quad+ (a_{11}+a_{12}\xi)\,f_0(a_{13}(a_{14}-\xi)) 
  \nonumber\\
  && \quad \quad \quad+ (a_{15}+a_{16}\xi)\,f_0(a_{17}(a_{18}-\xi))\,,
\label{eq:fit.P}
\end{eqnarray} 
where 
$\xi=\log(\rho/\textrm{g cm}^{-3})$, 
$\zeta = \log(p/\textrm{dyn}\,\textrm{cm}^{-2})$,
 $f_0(x) = \frac{1}{\mathrm{e}^x+1}$,
 and the coefficients $a_i$ for each case are tabulated in \cite{Haensel2004}.
The analytic approximations to SLY and FPS EoSs are shown in Figure~\ref{EOS_SLY_FPS_PLY}, as well as its first and second derivatives. We also include for comparison a simpler polytropic approximation, namely PLY, given by
\begin{equation}
\zeta=2\xi+5.29355\,.
\end{equation}
Despite the latter is not a realistic EoS apt to thoroughly represent NSs, it is  a {\it toy model} that will allow us to develop a detailed analysis of the derivatives of the EoSs and their crucial role in the calculation of stellar structure in $f(R)$ gravity.
%
\subsection{Numerical Results}
Solving the system of ordinary differential equations given by (\ref{dpdr})-(\ref{dmdr}) implies their integration from the centre of the NS to its surface, for which we assume boundary conditions: $m(r=0)=0$, $p(r=0)=p_c$, $\rho(r=0)=\rho_c$, $p(r=R)=10^{-12}p_c$ and $m(r=R)=M$. Once the solution is found, a couple of values $M$ and $R$ for the total mass and the radius, respectively, are established. Then, varying $\rho_{\rm c}$, a family of static configurations
$(M,R)$ is obtained. In order to perform the integration, we used a numerical code based on a fourth-order Runge-Kutta method with a variable step for the radial coordinate which is systematically shortened close to the NS surface, to account for rapid variations of the physical parameters in that region.  All metric functions are finite at $r=0$, thus ensuring that the obtained solutions are non-singular at the vicinity of the origin \cite{Henttunen2008}.\footnote{It is worth mentioning that there are spherically symmetric solutions to Palatini $R$-squared theory in which the $r=0$ region is not accesible. Such solutions represent wormholes \citep{Olmo2011b,Bambi2016,Olmo2016} (see \cite{Olmo2015} for wormholes generated by a one-parameter family of anisotropic fluids in the same theory).} 

\begin{figure}
\begin{center}
\resizebox{\hsize}{!}{
  \includegraphics[angle=-90]{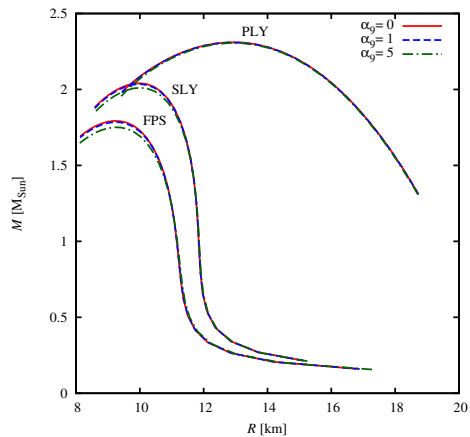}}
\caption{Static mass-radius configurations for the SLY, FPS and PLY EoSs, considering different values of the $\alpha$ parameter,  with the notation $\alpha_9=\alpha/10^9~{\rm cm}^2$. The GR case is recovered when $\alpha=0$, for which maximum masses are obtained in all cases.}
\label{MR}
\end{center}
\end{figure}

In Figure~\ref{MR}, we show the family of static configurations for the SLY, FPS and PLY EoSs, for three different values of the parameter $\alpha$, running from $\alpha=0$ (GR case) to $\alpha=5 \times 10^{9}$~cm$^2$. In this work we restrict the values of $\alpha$ in accordance to the constraints reported in \cite{Naf2010}.
 In all cases, the total mass corresponds to the value of the mass parameter at the surface of the star, where the internal metric coincides with the Schwarzschild solution. Although these constraints were obtained for $f(R)$ theories in the metric formalism, they represent a first attempt to study strong field scenarios in the Palatini formalism. While differences between modified gravity and GR are not appreciable for PLY EoS, significant changes can be noticed when the realistic EoSs are considered. Using the same initial conditions to compute the integration, the maximum of the configurations decreases when $\alpha$ increases. That is, in this particular choice of modified gravity NSs heavier than those in GR are not allowed for any of the two realistic EoSs. This feature of NSs in the $R+\alpha R^2$ model of modified gravity could be in tension with recent observations which evidence massive NSs, as the case of PSR~J1614-2230 ($M= (1.97 \pm 0.04)\ M_{\odot}$) \cite{Demorest2010} and PSR~J0348+0432 ($M=(2.01 \pm 0.04)\ M_{\odot}$) \cite{Antoniadis2013}. This tension may allow to place constraints on the parameter $\alpha$.

\begin{figure}
\begin{center}
\resizebox{\hsize}{!}{
  \includegraphics[angle=-90]{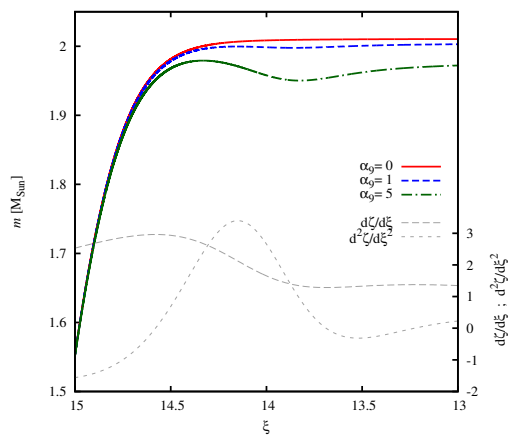}}
\caption{Mass parameter profile for the transition region between the core and the inner crust using the SLY EoS and for $\rho_c=4 \times 10^{15}$~gr~cm$^{-3}$. The counter-intuitive behaviour ${\rm d}m/{\rm d}\xi<0$ is more noticeable as $\alpha$ increases. First and second derivatives of the SLY EoS are plotted in dashed-gray lines to evidence the strong correlations of the mass profiles with the second derivative of the EoS, which becomes more prominent in the crust-core transition region ($\xi \sim$~14.0--14.3).}
\label{PERF}
\end{center}
\end{figure}

More remarkable features can be observed in Figure~\ref{PERF}, which shows the internal mass profiles obtained assuming a SLY EoS for $\rho_c=4 \times 10^{15}\ {\rm gr\ cm}^{-3}$ and three different values of the $\alpha$ parameter, together with the derivatives of the EoS used in grey-dashed lines.
 There exist internal regions where the mass parameter decreases with $\rho$, that is ${\rm d}m/{\rm d}\rho <0$, and, as the density and the pressure monotonously decrease with the radial coordinate, in those regions the mass parameter also decreases with the radius (${\rm d}m/{\rm d}r <0$). This unexpected behaviour becomes more noticeable when $\alpha$ increases. Moreover, the regions for which these features are observed are clearly correlated with those intervals in which the second derivative of the EoS becomes important, that is, close to phase-transition regions, in particular, in the crust-core transition, around $\xi = 14.1$. This correlation is also evidenced by the fact that ${\rm d}m/{\rm d}\rho$ remains always positive when the simpler PLY EoS with trivial derivatives is used. This counter-intuitive feature was previously reported in \cite{Barausse2008a} using the FPS EoS, where the authors claim that similar problems will appear in any theory involving higher derivatives in the matter fields than in the metric, since in such theories  the cumulative dependence of the metric on the matter field is not guaranteed.
 In fact, similar features on internal profiles of NSs were pointed out in \cite{Orellana2013}, where stellar structure was computed using a perturbative approach to find an approximated solution of the fourth-order differential equation system derived in the metric formalism. 

 This scenario motivates a careful analysis of the numerical approximation to realistic EoSs, since EoSs are generally constructed to successfully generate NSs in GR. Such a procedure could lead to the loss of important information contained in the first and second derivatives of the EoS, which are relevant in $f(R)$-gravity.  
 Thus, it is worth investigating possible descriptions of realistic NSs in a way such that the derivatives of the EoSs remain under control. We present in the next section a first approach to this problem, by proposing an alternative parametrisation to mimic the behaviour of EoSs close to phase-transition regions. We shall see that this parametrisation offers a simplified description of the nuclear matter on the crust and the core of NSs, allowing us to play with different values of ${\rm d}p/{\rm d}\rho$ and ${\rm d}^2p/{\rm d}\rho^2$ in order to re-interpret the results presented above. 

%
\section{A parametrisation for EOSs}

The interior of NSs is usually well-described by three distinct regions, namely the core, the inner crust and the outer crust (see Figure~\ref{EOS_SLY_FPS_PLY}). 
Each region can be roughly represented by a polytropic EoS with a characteristic polytropic index $\Gamma={\rm d}\log p/{\rm d}\log \rho={\rm d}\zeta/{\rm d}\xi$ \cite{Read2009}. 
We shall restrict our analysis to the densities which cover the core and the inner crust, as well as the corresponding transition region between them ($\xi \sim 14.2$), where the mass profile presents the most significant differences with respect to the GR case (see Figure~\ref{PERF}). 

\begin{figure}
  \begin{center}
    \resizebox{\hsize}{!}{
      \includegraphics[angle=-90]{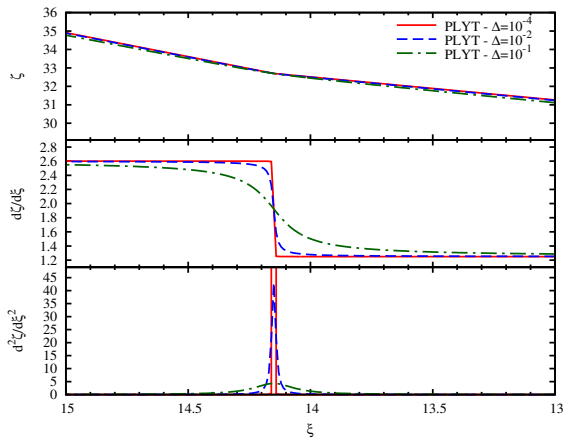}}
    \caption{PLYT EoSs around the crust-core transition $\xi \sim 14.1$ ($\xi \sim 11.9 $). In order to mimic SLY EoS within the range we are interested in, $13 \leq \xi \leq 15$, we set $\Gamma_2=2.6$, $\Gamma_1=1.25$, $\xi_0=14.15$ and $\zeta(\xi_0=14.15)=32.7$.  We use $\Delta=10^{-1}, 10^{-2}, 10^{-4}$ to illustrate significant examples.
    }
    \label{EOS_PLYT}
  \end{center}
\end{figure}

We choose to represent the above description by an arbitrary EoS, namely PLYT, shown in Figure~\ref{EOS_PLYT}, together with its first and second derivatives. In this representation, the first and second logarithmic derivatives of the PLYT EoS are given by
\begin{equation}
 \label{dzeta}
\frac{{\rm d}\zeta}{{\rm d}\xi} = \frac{\Gamma_2-\Gamma_1}{\pi}
\tan^{-1} \left(\frac{\xi-\xi_0}{\Delta}\right)+\frac{\Gamma_1+\Gamma_2}{2}\,,
\end{equation}
\begin{equation}
  \label{ddzeta}
\frac{{\rm d}^2\zeta}{{\rm d}\xi^2} =
\frac{\Gamma_2-\Gamma_1}{\pi}\frac{\Delta}{\Delta^2+\left(\xi-\xi_0\right)^2}\,,
\end{equation}
where $\Gamma_2$ and $\Gamma_1$ are the polytropic indices for the core and the inner crust, respectively. The parameter $\Delta$ characterises the width of the transition region, allowing us to control there the first and second derivatives of the PLYT EoS. Since second derivatives are unconstrained by thermodynamics, $\Delta$ can be chosen as small as desired.\footnote{See \cite{Bejger2005} for a discussion about astrophysical scenarios of the formation of a mixed-phase core in neutron stars.}
In the limiting case in which both polytropic EoSs are matched with a hard phase-transition between the core and the inner crust, i.e. $\Delta$ near to but different from 0. The explicit form of PLYT EoS can be obtained by integrating Eqn.~(\ref{dzeta}).

\begin{figure}
  \begin{center}
    \resizebox{\hsize}{!}{
      \includegraphics[angle=-90]{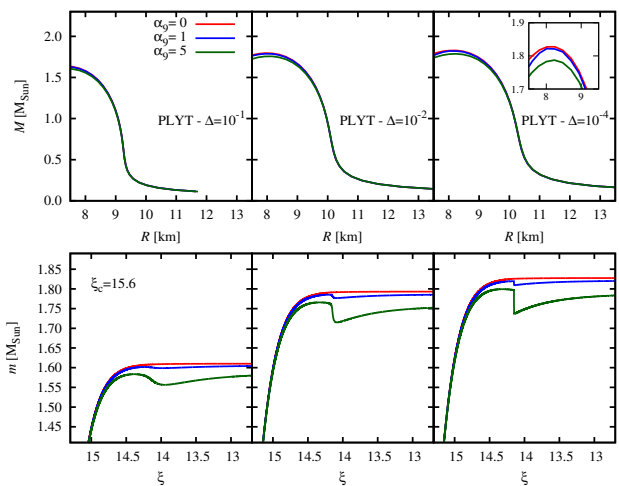}}
    \caption{Static mass-radius configurations (upper panels) and mass parameter profiles (lower panels) for the PLYT EoS, considering different values of the $\alpha$ parameter, and, from left to right, for three different values of the width of the crust-core transition region ($\Delta=10^{-1}$, $10^{-2}$ and $10^{-4}$). The GR case is recovered when $\alpha=0$. Mass parameter profiles correspond to $\rho_c=4 \times 10^{15}$~gr~cm$^{-3}$. The counter-intuitive behaviour ${\rm d}m/{\rm d}\xi<0$ occurs for the three values of $\Delta$ in the crust-core transition region ($\xi = 14.15$). }
    \label{MR_PERF_PLYT}
  \end{center}
\end{figure}

The mass-radius configurations as well as the internal profiles obtained for PLYT EoS are shown in Figure~\ref{MR_PERF_PLYT}. In order to mimic the SLY EoS within the range we are interested in, $13 \leq \xi \leq 15$, we set $\Gamma_2=2.6$, $\Gamma_1=1.25$, $\xi_0=14.15$ and $\zeta_0(\xi_0=14.15)=32.7$ in Eqns.~(\ref{dzeta}) and (\ref{ddzeta}).  We use $\Delta=10^{-1}, 10^{-2}$ as examples that produce similar results to those obtained with the SLY EoS in the previous Section. Mass-radius configurations reach lower maximum masses when $\alpha$ is increased, and the peculiar behaviour ${\rm d}m/{\rm d}\xi< 0$ can still be observed in the internal profiles, being more pronounced for smaller $\Delta$. 
 Although the limiting case $\Delta = 0$ cannot be analysed in the present formalism due to the discontinuity in the first derivative of the EoS, 
 the results shown in
 Figure~\ref{MR_PERF_PLYT}
 suggest that the sequence of decreasing values of $\Delta$ ultimate leads to a discontinuity in $m(r)$. 
This behaviour indicates that, even if phase transitions in the EoS are allowed, the undesirable behaviour ${\rm d}m/{\rm d}\xi< 0$ in the mass profile will not removed.  
Furthermore, mass-radius configurations do not allow heavier NSs than in GR for any plausible $\alpha > 0$ in this theory.
%
\section{Discussion}
In order to investigate whether $f(R)$-theories in the Palatini formalism can be used to describe astrophysical scenarios in the strong curvature regime, we studied the internal structure of NSs in the theory defined by $f(R)=R+\alpha R^2$. In contrast to the metric formalism, the modified TOV equations have derivatives of the metric up to the second order, as in the GR case. However, in spite of this advantage, the integration involves some extra considerations since derivatives of the EoS are present in the structure equations. 

Considering the SLY EoS commonly used to compute NSs, we obtained results consistent with previous studies (in which the FPS EoS was used \cite{Barausse2008a}) regarding the static mass-radius configurations and internal mass profiles.
 Concerning the mass-radius relations, lower maximum masses than those in the GR case are obtained, although the differences are not large enough to fully constrain the parameter $\alpha$ by observational evidence of the most massive NSs. A more serious problem is found when the internal structure of these models is analysed. A counter-intuitive behaviour is observed in the mass profiles, which include regions where ${\rm d}m/{\rm d}\rho<0$. It was claimed in \cite{Barausse2008a} that this feature is a natural consequence of theories of gravity involving higher order derivatives in the matter fields than in the metric. Assuming the validity of the realistic EoS, it may be possible to limit the parameter $\alpha$ to values lower than $10^9$cm$^2$ if ${\rm d}m/{\rm d}r>0$ is required all through the interior of the star. However, EoSs for matter in the extremely high density regime are usually constrained by fitting the structure of NSs in GR, where only the zeroth-order relation between $\rho$ and $p$ is relevant. Then, it seems inappropriate to use an EoS to constrain alternative theories of gravity without imposing the bias $\alpha=0$. This is of course an intricate problem because NSs are actually the only natural laboratories where properties of high density matter can be tested.  

Thus, in this work we also studied an alternative parametrisation for the EoSs, namely PLYT, that simply accounts for the core and the crust regions of the NS. This is achieved by means of polytropic relations connected continuously and analytically, which mimic the phase-transition between both regions. This new parametrisation of the EoS allows us to control its first and second derivatives. The trends of mass-radius configurations found using an analytic approximation to realistic SLY and FPS EoSs are recovered, as well as their internal profiles. We found that even in the limiting case representing a hard phase transition between the core and the crust of the compact star, the peculiar behaviour of the mass parameter profile is unavoidable, and lighter NSs than those calculated with GR are obtained. 
Our results also indicate that in the limit $\Delta = 0$, there will be a singularity in the curvature, due to the
discontinuity in $m(r)$. 
These features seem to suggest that the problems claimed to be characteristic of NSs in Palatini $f(R)$ theories are indeed rooted to the nature of the field equations, and core-crust phase transitions in EoSs are not capable to counteract this dependence.

 To conclude, we would like to mention two lines of research that are a natural extension of this work. The first one is the possible existence of wormhole-like solutions that may arise from particular choices of the EoS and the function $f(R)$. The second is the study of the stability of the calculated NSs, which would be very important to ensure that configurations using different parametrisations of the EoS can be realised under $R$-squared gravity. Such studies are left for future work.

%
\begin{acknowledgements}
FATP, FG, GER and MO acknowledge support from CONICET. 
 FATP would like to acknowledge support from Programa de Doutorado Cooperativo CLAF/ICTP. 
SEPB acknowledges support from FAPERJ and UERJ.
\end{acknowledgements}

\bibliographystyle{apsrev} 
\bibliography{bibliography}   

\end{document}